      [1999/12/01 v1.4c Il Nuovo Cimento]
\documentclass{cimento}

\usepackage{graphicx}  
\title{GRB Progenitors and Environment}
\author{Davide Lazzati\from{ins:x}}
\instlist{\inst{ins:x} JILA, 440 UCB, University of 
Colorado, Boulder, CO 80309-0440, USA}
\PACSes{\PACSit{98.76.Rz - 98.38.j - 95.30.Dr}{}}
\begin{document}

\maketitle

\begin{abstract}
The study and knowledge of the environment of Gamma-Ray Bursts is of
great interest from many points of view.  For high redshift ($z>0.5$)
events, the structure of the ambient medium is one of the best
indicators of the nature and properties of the progenitor. It also
tells us about the last stages of the pre-explosion evolution of the
progenitor. In addition, it is interesting in its own as a sample of
the interstellar medium in a high redshift galaxy. Measures of the
density and structure of the GRB environment are however sparse, and
different methods yield different, often incompatible, results. I will
review the methods and results with particular emphasis on the case of
GRB~021004, a puzzling but highly informative event. I will finally
underline the advancements that will be possible in the Swift era.
\end{abstract}

\section{Introduction}

The medium surrounding the location where a Gamma-Ray Burst (GRB)
exploded has become an issue of study since the detection of the first
afterglows. In fact, it is thanks to the absorption features imprinted
by the ambient medium on the afterglow spectrum of GRB~970507 that the
first GRB redshift was ever measured~\cite{ref:met97}. Absorption
measurements, however, usually allow us only to infer the column
density of the material lying along the line of sight to the observer
and not its structure and geometry.

Modeling of the afterglow light curve is a tool that can be used to
infer the environment structure in the surroundings of the GRB, up to
a distance of several tens of parsec to several parsec, i.e. the
distance traveled by the relativistic fireball before becoming
sub-relativistic. The largely heuristic nature of the afterglow theory
prevents us from drawing consistent results from the
method. Alternative methodologies involve either reverberation
techniques or time variability of the opacity of the surrounding
medium to constrain certain properties of the environment.

In this contribution I will review the techniques used to measure the
environment properties of cosmological GRBs and discuss the results
obtained. I will focus on the discrepancy in the results of different
approaches and discuss what Swift, with its rapid follow up, will
enable us to derive.

\section{Theory}

The properties of the GRB close environment heavily depend on the
object that is assumed to be the burst progenitor. It is now widely
accepted that long GRBs are associated to the explosions of massive
stars~\cite{ref:gal98,ref:sta03,ref:hjo03}. More uncertain, on the
other hand, is the nature of the short burst progenitor. This is due
to the lack of afterglow observations for short GRBs. 

\subsection{Wind profile}

The association of long GRBs with massive star progenitors calls for a
structured environment, at least in the close vicinity to the
explosion site. Massive stars, and especially the Wolf-Rayet (WR)
stars supposed to be progenitors of long GRBs, produce heavy and fast
winds, which generate a decaying density profile of the form:
\begin{equation}
n=3\times10^{13}\,\dot{M}_{-5}\, R_{11}^{-2}\,v_8^{-1} \quad
\rm{cm}^{-3}
\end{equation}
where\footnote{Here and in the following we will adopt the notation
$Q_x=10^x\,Q$ and will use cgs units, unless specified.}
$\dot{M}_{-5}$ is the mass loss rate in units of $10^{-5}$ solar
masses per year and $v_8$ the wind terminal velocity in units of
$10^8$~cm~s$^{-1}$. Such profile is however only a first approximation
to the environment of the massive stars. Several complications still
awaits to be fully understood:
\begin{itemize}
\item the mass ejection rate of a star less than $\sim100$~years to 
explosion may be far from constant, and so big deviations from the
$R^{-2}$ profile could be present;
\item the wind of WR stars is observed to be 
clumped~\cite{ref:sch04,ref:mar03};
\item the interaction of the wind with the surrounding material generates 
a shock structure at radii of the order of parsecs.
\end{itemize}

\begin{figure}
\centerline{\includegraphics[width=0.8\textwidth]{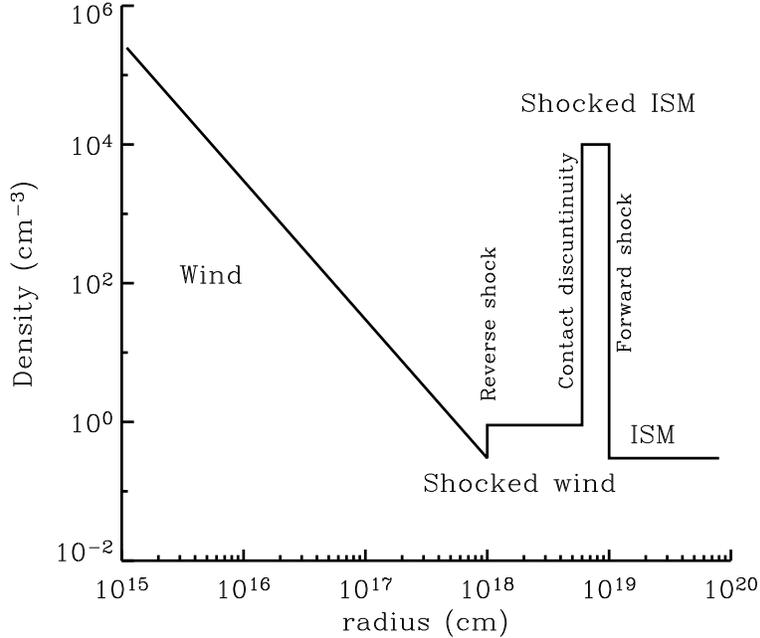}}
\caption{Schematic view of the density structure in a simple case 
of wind-ISM interaction~\cite{ref:che04}.
\label{fig:wind}}
\end{figure}

Some work has been devoted to the last aspect, since the transition
from a wind-dominated environment to the shock-shaped one and
eventually to the molecular cloud should be at distances from the
progenitor where the afterglow radiation is produced. In the simplest
case, the transition should consist of a contact discontinuity between
the wind material and the interstellar medium (ISM) and by two shocks
(see Fig.~\ref{fig:wind}). A forward shock propagates out in the ISM
while a reverse shock propagates backward in the Wind, creating a
region of hot wind material with uniform
density~\cite{ref:che04}. This uniform region is interesting for
observations (see below). Its extent may very dramatically depending
on the wind and ISM properties and, in some cases of high wind power,
the shock can bounce back on the star wind and create a wide uniform
hot region~\cite{ref:dwa03}. Going from 1-D to 2-D and 3-D models adds
the complication of turbulence and clumping to the
picture~\cite{ref:vantc}.

Another important aspect of wind theory that is not entirely clear is
where from the wind starts. In other words, should the wind start
immediately at the surface of the star, the Thompson opacity would be
\begin{equation}
\tau_T=\frac{2}{R_{11}} \, \dot{M}_{-5}\,v_8^{-1}
\end{equation}

Since the radius of the progenitor star is constrained by simulations
to be $R_\star\le10^{11}$~cm~\cite{ref:zha04}, for fiducial mass loss
histories and wind speeds, an optically thick environment would be
expected. This would create strong effects due to the interaction of
the primary photons with a dense medium.

\subsection{Uniform ISM}

More uniform interstellar media are favored by different classes of
progenitors. The main alternative to massive stars, and leading
candidate for short GRB progenitor, is the merger of a binary system
made of two compact objects. Such binary systems are long lived and,
due to SN kicks, they travel outside their formation site and possibly
out to the intergalactic medium. The fact that most long GRB afterglow
observations favor this scenario is still a not entirely understood
aspect of the problem.

\subsection{Photon-ISM interaction: pairs}

An important effect of the propagation of $\gamma$-ray photons in the
ambient medium is that a fraction of these photons are
Thompson-scattered by the electrons. Scattered photons are ideal
target for photon-photon interactions to give electron-positron pairs:
\begin{equation}
\gamma+\gamma=e^++e^-
\end{equation}
The problem can be dealt with if the ambient medium is optically
thin~\cite{ref:bel02}. The main consequence of the pair enrichment is
that in the early afterglow the typical synchrotron frequency is small
(since the same energy is divided into many more leptons). The early
afterglow should therefore be optically dominated\cite{ref:bel05}
(rather than X-ray).

\subsection{Photon-ISM interaction: dust}

At later times and larger radii ($r>10^{16}$~cm), the energy and flux
of photons is not large enough to produce pairs. Still, the photons
interact with the environment and modify it. Let us first consider
dust particles. Dust grains absorb photons mostly in the UV and soft
X-ray band. Absorption results mostly in heating of the grain. If the
heating is larger than the cooling rate (which is mostly radiative),
the internal agitation of molecules becomes larger than the binding
energy and the grain evaporates~\cite{ref:wax00,ref:per02}. Higher
energy photons, instead, ionize the ion until a surface potential
larger than the binding energy is created. At this point the ion may
either break into pieces~\cite{ref:fru01}, in a runaway dissociation,
or eject charged ions to re-establish equilibrium~\cite{ref:per02}. In
the first case, the dust destruction can be effective to large radii,
out to tens of parsec. In the second, more conservative, estimate, the
dust distribution is affected out to large radii, but dust grains are
completely destroyed only at smaller radii (out to $\sim10$~pc at
most).

\subsection{Photon-ISM interaction: ions}

Besides the interaction with dust grains, photons strongly interact
with atoms and ions. The main effect of this interaction is the
photoionization of the elements in the gaseous phase. What makes this
process interesting to GRB environment studies (as well as the dust
interaction discussed above) is that the photoionization is observed
``live on stage'', and as a consequence the column density of absorber
can be observed to decrease in
time~\cite{ref:per98,ref:boe99,ref:laz01,ref:laz02,ref:laz03}. Since
the timescale with which the species are ionized grows with the square
of distance, the timescale with which the column density decreases
give us constraints on the distance of the absorber. Coupling this
with the early time column density we can derive an average density as
well.

\subsection{Reverberation}

It is in principle possible to derive the overall structure and
density of the ambient medium by studying the reverberation of primary
photons scattered or reprocessed by ions. These studies were triggered
by the tentative detection of X-ray lines in the early afterglow of
GRBs~\cite{ref:pir99,ref:ree02,ref:ghi02}. Unfortunately, such
detections have never been confirmed, and so the results on the
environment~\cite{ref:laz99,ref:laz02b} (very high densities in the
close vicinity of the burster) are still to be considerate tentative.

\section{Observations}

In this section I will summarize the observational results that have
been obtained by applying the techniques described above to afterglow
observations.

\subsection{Afterglow modeling}

The first estimates on the density and radial structure of the GRB
ambient medium were derived by multi-wavelength and multi-epoch
modeling of the afterglow intensity~\cite{ref:wij99}. The afterglow
model has four main unknowns: the density of the ISM, the energy of
the fireball and the two shock parameters $\epsilon_e$ and
$\epsilon_B$. On the other hand, there are four observables: the three
break frequencies (self-absorption, peak and cooling) and the
normalization. For this reason the system of equations can be inverted
to derive a solution. Generally, these analyses~\cite{ref:pan02} yield
preferentially uniform low-density environments, in striking contrast
with the prediction of the collapsar progenitor model~\cite{ref:che99}.

The discrepancy could be due to oversimplification in the afterglow
model~\cite{ref:ros03} or to neglecting the effect of long lasting
energy input from the inner engine~\cite{ref:bjo02}. Future
observation will hopefully clarify this issue.

\subsection{X-ray absorption}

\begin{figure}
\centerline{\includegraphics[width=0.6\textwidth]{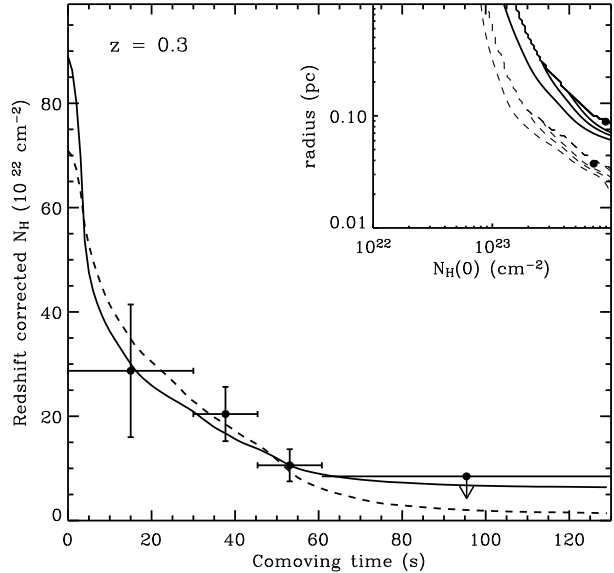}}
\caption{Column density measurements for GRB~000528~\cite{ref:fro04}.
The solid and dashed lines are models of column density evolution for
an uniform cloud and a shell surrounding the burst site,
respectively. The inset shows the $\chi^2$ contours and best fit
parameters for the two cases.
\label{fig:xcon}}
\end{figure}

Several constraints have been obtained by modeling time-dependent
X-ray absorption detected during the prompt phase of {\it{Beppo}}SAX
GRBs. In some bursts the absorption in the soft X-ray continuum is
observed to decrease in time, as predicted by the progressive
photoionization of the environment (see Fig.~\ref{fig:xcon}). Even
though the data are not of sufficient quality to derive the radial
structure of the absorber, the observations can be reproduced only if
a high density compact absorber surrounds the GRB
site\cite{ref:laz02,ref:fro04} ($n\sim10^{6}$~cm$^{-3}$;
$r\sim0.1$~pc).

In two other bursts, instead, a narrow absorption feature was observed
to weaken in time. The best case is that of
GRB~990705~\cite{ref:ama00}. A narrow absorption feature corresponding
to the fully ionized $K_\alpha$ absorption edge of iron is detected
only in the first $\sim10$ seconds of the observation. The inferred
amount of iron required to produce such a feature is huge (more than
100 solar masses). If the feature is considered to be due to resonant
scattering from outward moving iron~\cite{ref:laz01a}, a more
reasonable amount of iron is required. Again, this detection points to
a high density $n\ge10^{10}$~cm$^{-3}$ in close vicinity to the burst
($r\sim0.01$~pc). An analogous feature was detected in
GRB~011211~\cite{ref:fro04a}, yielding similar constraints, even
though a faster outflow of the absorber was implied.

\subsection{Optical resonant lines \& GRB~021004}

\begin{figure}
\centerline{\includegraphics[width=0.6\textwidth,angle=-90]{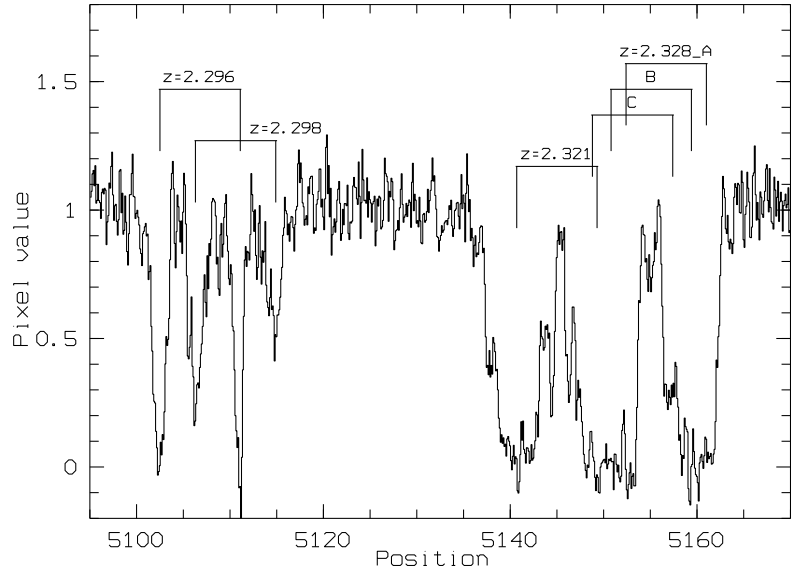}}
\caption{The complex CIV absorption system in the spectrum of 
GRB~021004~\cite{ref:fio05}.
\label{fig:021004}}
\end{figure}

\begin{figure}[t]
\centerline{\includegraphics[width=0.35\textwidth,angle=90]{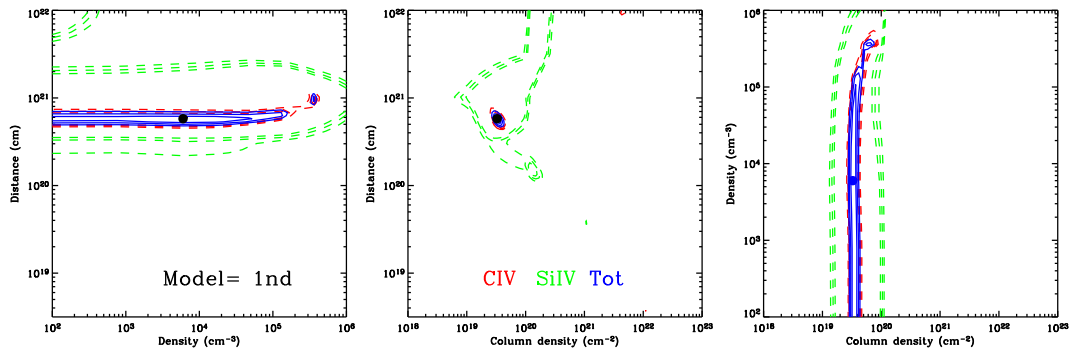}}
\caption{$\chi^2$ contours in the density, column density and distance 
planes for the evolution of the column density if CIV and SiIV (dashed
lines) in the spectra of GRB~021004. The solid line shows the combined
contours. Levels are for 1, 2 and 3 standard deviations.
\label{fig:021004a}}
\end{figure}

A more recent development has been possible thanks to the rapid and
extended spectroscopic follow-up of GRB~021004. This GRB had a complex
absorption system in its optical spectrum, characterized by the
presence of multiple absorbers with different velocities, spanning a
range of $\sim3000$~km~s$^{-1}$ in proper speed (see
Fig.~\ref{fig:021004}). The absorption systems were observed
repeatedly in time from few hours to several days after the burst
expolsion. This allows to derive the time evolution of the absorption
for two elements: CIV and SiIV. Adopting a time dependent
photoionization and dust destruction code~\cite{ref:per02} it is
possible to model the evolution (or non-evolution) and derive limits
on the distance and density of the various absorption systems. The
results of this modelling are shown in Fig.~\ref{fig:021004a}. The
absence of a strong evolution of the equivalent widths of the lines
with time requires a large distance for the absorber, of the order of
tens of parsecs\footnote{The actual best fit density depends on the
model spectrum for the prompt $\gamma$-ray and optical flash
emission. The figure shows a model with small optical emission.}. The
density, however, is not well constrained~\cite{ref:fla05}.

\section{Summary and Discussion}

\begin{figure}[t]
\centerline{\includegraphics[width=0.8\textwidth,height=0.5\textwidth]
{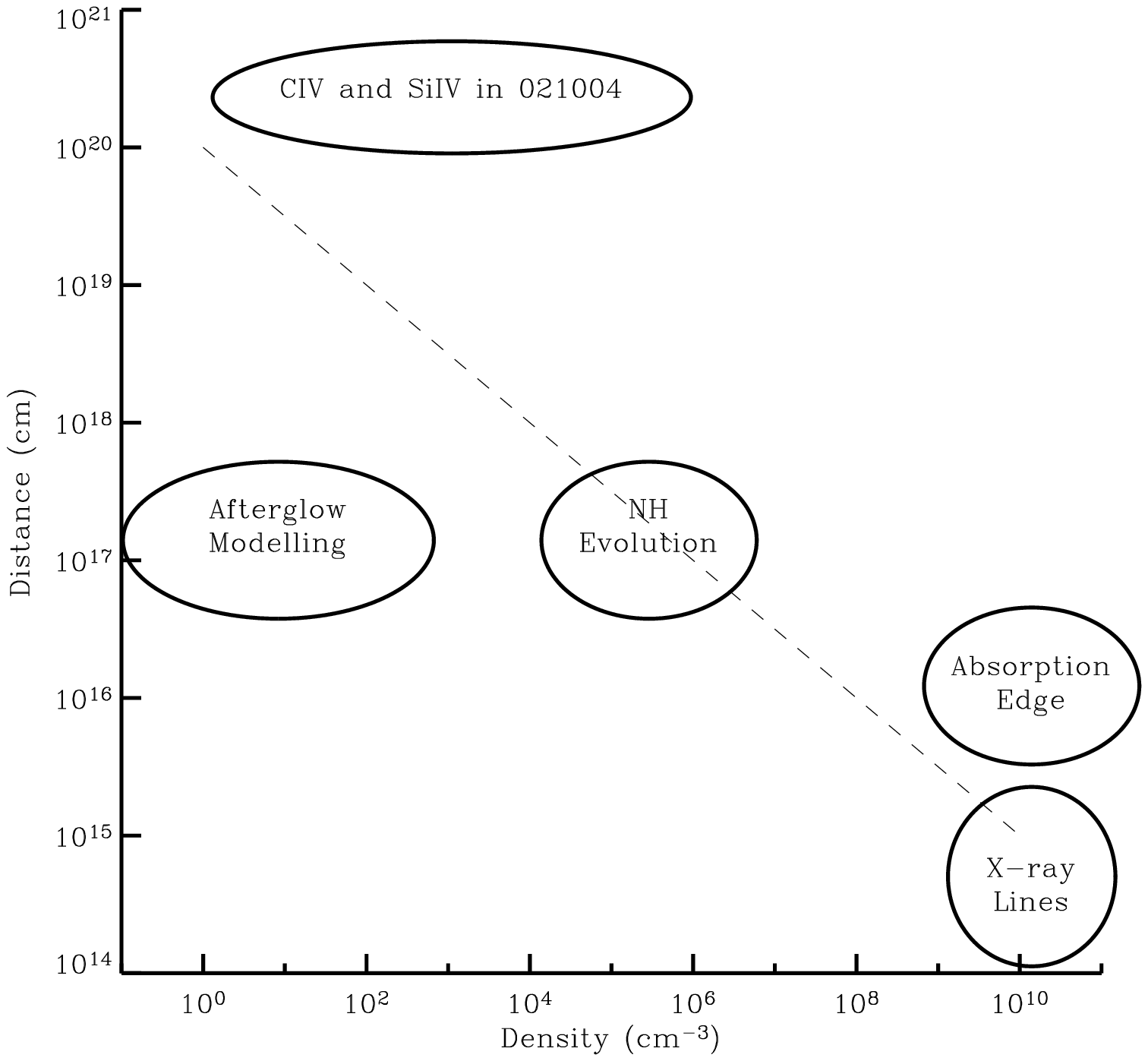}}
\caption{Summary cartoon of the average value of environment 
density measures performed with different techniques (and usually on
different GRBs). The dashed line represents a wind profile.
\label{fig:summ}}
\end{figure}

Figure~\ref{fig:summ} shows a graphic summary of all the available
measurements of density of the GRB ambient medium, performed with
different techniques that are sensitive at different distances. The
size of the ellipses roughly represents the scatter of the measures
for each method. While it is true that different techniques have been
applied to different events, still the scatter is considerable and the
agreement totally absent. The fact that X-ray based measurements are
offset to higher densities does not surprises, since only if the
(column) density is large something can be actually measured by its
X-ray opacity. However, ten orders of magnitudes of difference in
density with respect to afterglow modelling are quite big. The
measurement performed in GRB~021004~\cite{ref:fla05} is roughly
consistent with the afterglow density measurements. Unfortunately the
high velocity of the lines calls for a wind profile, while consistency
would be reached only in the case of a uniform medium.

The bottom line of Fig.~\ref{fig:summ} is that what we still do not
have a clear observational picture of what the circum-burst material
looks like. Some observations may be circumstantial and some theory
may need refinement. 

In the Swift era this situation may change. Early observations will
allow us to study in more detail the initial stages of the
fireball-ISM interaction. This should allow us to better understand
the physics of the reverse/forward shock and the importance of
neutrons and pair enrichment. In addition, early observations are
mandatory to study the clumpiness of the medium, since in the early
stages only a small portion of the fireball is visible and therefore
local properties of the ISM can be studied. Early photometry will also
allow us to detect, if any, signatures of the dense environment close
to a massive star, if the wind profile continues to the stellar
boundary. Swift will also have a large impact on ground-based
observations. Since photon-ISM interactions are stronger and faster at
early times, prompt Swift localization will enablke us to study any
time dependence in the opacity of the ISM. As discussed above, these
are a powerful tool to study the radial profile of the absorber.

\acknowledgments
I'd like to thank Fabrizio Fiore, Gabriele Ghisellini, Rosalba Perna
and Joe Flasher for useful collaborations which led to many of the
results presented in this talk. This work was supported in part by NSF
grant AST-0307502 and NASA Astrophysical Theory Grant NAG5-12035.

\end{document}